\documentclass[aps,prc,twocolumn,superscriptaddress,showpacs]{revtex4}

\usepackage{graphicx}

\begin{document}

\title{Comment on ``$^{187}$Re($\gamma$,n) cross section close to and above
the neutron threshold''}

\author{Thomas Rauscher}

\affiliation{Departement f\"ur Physik und Astronomie, Universit\"at Basel, 
CH-4056 Basel, Switzerland}


\date{\today}

\begin{abstract}
The work of M\"uller et al.\ [Phys.\ Rev.\ C {\bf 73}, 025804 (2006)]
provides interesting experimental data on neutron emission by
photodisintegration of $^{187}$Re. However,
the comparison to theory and the discussed implications for the Re/Os clock
require considerable amendment.
\end{abstract}

\pacs{26.20.+f, 25.40.Lw, 25.20.Dc, 27.70.+q}

\maketitle

The recent paper by M\"uller, Kretschmer, Sonnabend, Zilges, and Galaviz
\cite{mkszg} (hereafter referred to as MKSZG) presents
measurements of the neutron emission from $^{187}$Re induced by Bremsstrahlung
photons. Derived cross sections are compared to predictions from
two different Hauser-Feshbach models, NON-SMOKER \cite{rath,rauphoto} and
MOST \cite{gorkhan,aik}, showing good agreement.
In the astrophysical application, the relevant quantity is the
neutron capture on $^{186}$Re.
Due to the short half-life of $^{186}$Re
($T_{1/2}=3.7$ d), a direct measurement of the capture cross section is
difficult. Therefore, MKSZG attempt to extract information on the reliability
of the predictions from their data. While the data from the nice
measurement are accurate and interesting, this latter interpretation and
comparison to theory necessitates some remarks.

First, an apparent confusion in the comparison between the recommended
value of Bao et al.\ \cite{bao00} and the two models has to be rectified.
In fact,
the stellar enhancement factors and the MACS for $^{186}$Re(n,$\gamma$)
given in \cite{bao00} are
those of the NON-SMOKER code, as published in \cite{rath}. There are no
experimental data included in \cite{bao00} for this nucleus. The
recommended value is derived by a renormalization of the NON-SMOKER value
with a factor accounting for systematic errors in the NON-SMOKER
calculations. The procedure is explained in detail
in Bao et al.\ \cite{bao00} (see Eqs.\ 1, 2 in that paper). The error bars
are also derived in that procedure. Systematic errors may arise because
both models, NON-SMOKER and MOST, are global models which do not include
local data or parameterizations of the relevant nuclear properties but
rather try to make predictions based on either microscopic models or
global parameterizations. However, a comparison of the unrenormalized
NON-SMOKER value of 1365 mb as quoted in \cite{bao00} (it can also be
extracted from \cite{rath,rath2}) to the recommended value of 1550$\pm$250 mb
shows that there is no large systematic
error for nuclei with neutron numbers around $N\simeq 111$, according to the
applied procedure. The quoted error bar is a very conservative estimate.
Nevertheless, the accuracy at this neutron number seems to be better
than the average uncertainty of 30\% found earlier \cite{rtk}.

Considering the above, the fact that the ``MACS calculated with the updated
NON-SMOKER code is close to the value'' given in Bao et al.\ \cite{bao00},
as stated by MKSZG, is not surprising. However, the statement ``Both models
predict a (n,$\gamma$) cross section which is smaller than the value
recommended'' in \cite{bao00} (quotation from MKSZG) clearly is unjustified.
Firstly, the ``updated'' NON-SMOKER value clearly lies within the error bar
of the recommended value. Secondly, even if this were not the case,
this would not
allow to draw conclusions of the kind MKSZG present because the
Bao et al.\ value is the same theoretical value but including corrections
for systematic errors, based on the same theory. More accurately,
a renormalization and
an error bar of similar
magnitude should be
applied to the quoted values of NON-SMOKER and MOST when attempting
a comparison to the Bao et al.\ value \cite{bao00}. This would reveal that the
NON-SMOKER and the Bao et al.\ value are actually the same (except for a 9\%
difference, see below).

Also the statement by MKSZG that the ``disagreement
between the two models may have its origin in the lack of precise nuclear
data that enter into both models'' is an inauspicious wording. As stated above, both
models are global models which deliberately refrain from using local data.
However,
it has to be agreed that experimental data is needed, to test the
global predictions and to improve the global parameterizations used in the
calculations. This can only be achieved through a detailed comparison of
theoretical and experimental results in test calculations.

One may notice that the ``updated'' NON-SMOKER MACS of 1485 mb given by MKSZG
in their Table III (and incorrectly referenced as \cite{rauphoto}; it should
rather be ``T. Rauscher, private communication'') is slightly larger than the
unrenormalized value quoted in \cite{bao00}.
This is due to the fact that the calculation performed for MKSZG included
E1, M1, and E2 photon transitions, whereas the older calculation 
\cite{rath,rath2,rauphoto} only
included E1 and M1. The 9\% difference shows that E2 transitions are
not very important but contribute to the final
cross section nevertheless.

MKSZG conclude that the ``fact that both model predictions are in good
agreement with our data but could not reduce the overproduction of $^{186}$Os
supports the idea that the adopted value of the $^{186}$Os(n,$\gamma$)
cross section is too small''. While the experiments at n\_TOF and at FZK
may indeed suggest this, a similar conclusion from the MKSZG data and
comparison to the Hauser-Feshbach predictions is premature without a detailed
study of the sensitivities of the predictions to the actually measured
quantities. For example, the ($\gamma$,n) experiment on $^{187}$Re in the
ground state tests different neutron- and $\gamma$-transitions than a
neutron capture measurement on $^{186}$Re. The photodisintegration includes
a single E1 $\gamma$ transition from the ground state of $^{187}$Re,
populating states with spins and parities of 3/2$^-$, 5/2$^-$, 7/2$^-$.
These states can then decay by neutron emission to ground and excited states
in $^{186}$Re. Thus, the properties of those low-lying states in $^{186}$Re and the
level density above the last considered excited state enter the calculated
cross section. The situation is different for the (n,$\gamma$) direction.
Assuming neutron capture on the ground state of $^{186}$Re (the small
stellar enhancement factor shows that capture on excited states contributes
little), s-wave neutron transitions populate states in $^{187}$Re with
$J^\pi=$1/2$^-$, 3/2$^-$. These states can then deexcite either by a single
E1 transition to the ground state of $^{187}$Re (from the 3/2$^-$ state
only) or via $\gamma$ cascades involving different kinds of transitions. Thus,
the properties of low-lying states and the level density in $^{187}$Re
enter the calculation of the cross section, as well as a contribution of E1
and M1 cascades (E2 transitions were not considered in the theoretical
values quoted for the photodisintegration). Apparently, at best the ($\gamma$,n)
experiment tests a subset of transitions relevant for (n,$\gamma$) plus some
transitions which are of minor importance for the capture. This is true
for most photon-induced experiments \cite{mohr04}.

Further inspection suggests that the differences in the energy dependence
of the ($\gamma$,n) cross section predicted in the two models are most
likely due to the different treatment of the nuclear level density. As can
be found in \cite{rath2}, 10 excited states in $^{186}$Re up to an
excitation energy of 314 keV are included in the calculation (the
photodisintegration cross sections provided in \cite{rauphoto} were
computed with the same model and inputs as given in \cite{rath2}). Both models,
NON-SMOKER and MOST, include these states. Above 314 keV excitation energy,
an average nuclear level density is invoked. The description of this level
density is different in the two models, leading to a different number of
possible transitions. Therefore, a difference in the calculated energy
dependence in the two models is expected for photon energies of more than
314 keV above the ($\gamma$,n) threshold whereas a very similar behavior is
to be expected below that energy. This is confirmed by Fig.\ 6 of MKSZG.
This also underlines that not only neutron emission to the ground state of
$^{186}$Re is important but that neutron transitions to excited states also
contribute. It should be kept in mind that the level density
probed here is the one in the odd-odd nucleus $^{186}$Re and not the one in
the odd-even $^{187}$Re which is relevant for the neutron capture
reaction.

Finally, Fig.\ 5 of MKSZG suggests that the energy dependence of the cross section
differs from the measured one. The top panel of the figure shows
that the energy dependence is compatible with s-wave neutron emission. (It is
misleading here and in Fig.\ 6 of MKSZG to call it an
``experimental cross section''. It is more appropriately called a ``cross
section derived under the assumption of pure s-wave neutron emission''.) Were
the energy dependence of the models compatible with the experiment, it
should be possible to fit the values
shown in the lower two panels with a constant normalization factor
$f$.
However, the independence from the
energy cut-off $E_\mathrm{max}$ of the integration in Eq.\ 6 of MKSZG is
only destroyed by the normalization factors obtained for the two lowest
cut-off energies. For these, the energy range close above the threshold
$S_\mathrm{n}$ contributes significantly to the integral.
Therefore these values of the
normalization factors are very sensitive to the threshold region and the
integration performed in the small energy interval just above the threshold.
In addition to the quoted possibilities of deviations in that region,
experimental uncertainties regarding the precise shape of the Bremsstrahlung
spectra and numerical problems in the theoretical calculation, two further
possible sources of deviations have to be mentioned. Close to channel
openings, width fluctuations
can occur in the cross section which have to be
accounted for by a correction to the calculation \cite{rath}. It remains to be
checked whether such fluctuations were treated
appropriately. Secondly, the mesh used for the integration might
have been simply too coarse as the determination of the renormalization
factors for the theoretical models relies
on the set of cross sections provided to
MKSZG and the mesh on which the spectral energy
distribution $N_\gamma$ is evaluated. Considering these additional
uncertainties, it is premature to conclude that the theoretical
values show a different energy behavior than found in the experiment
and to imply possible implications on the reliability of
the related (n,$\gamma$) cross sections.

Summarizing, it has to be emphasized that an interesting, accurate
measurement has been presented by MKSZG.
This comment is intended
as a correction of and an amendment to the discussion
included in MKSZG.

\end{document}